\documentclass[a4paper,12pt]{article}
\usepackage{amsfonts}
\usepackage{latexsym}
\usepackage{amssymb}
\begin{document}
\date{}

\title{Exploitation of the Dissipation Inequality in\\General Relativistic
Continuum Thermodynamics 
\thanks{Dedicated to G\'{e}rard A. Maugin on the occasion of his
70th birthday}}
\author{W. Muschik\footnote{Corresponding author: muschik@physik.tu-berlin.de}
\quad
and\quad H.-H. v. Borzeszkowski\footnote{borzeszk@itp.physik.tu-berlin.de,
borzeszk@t-online.de}\\
Institut f\"ur Theoretische Physik\\
Technische Universit\"at Berlin\\
Hardenbergstr. 36\\D-10623 BERLIN,  Germany}
\maketitle

            \newcommand{\be}{\begin{equation}}
            \newcommand{\beg}[1]{\begin{equation}\label{#1}}
            \newcommand{\ee}{\end{equation}\normalsize}
            \newcommand{\bee}[1]{\begin{equation}\label{#1}}
            \newcommand{\bey}{\begin{eqnarray}}
            \newcommand{\byy}[1]{\begin{eqnarray}\label{#1}}
            \newcommand{\eey}{\end{eqnarray}\normalsize}
            \newcommand{\beo}{\begin{eqnarray}\normalsize}
            \newcommand{\R}[1]{(\ref{#1})}
            \newcommand{\C}[1]{\cite{#1}}

            \newcommand{\mvec}[1]{\mbox{\boldmath{$#1$}}}
            \newcommand{\x}{(\!\mvec{x}, t)}
            \newcommand{\m}{\mvec{m}}
            \newcommand{\F}{{\cal F}}
            \newcommand{\n}{\mvec{n}}
            \newcommand{\argm}{(\m ,\mvec{x}, t)}
            \newcommand{\argn}{(\n ,\mvec{x}, t)}
            \newcommand{\T}[1]{\widetilde{#1}}
            \newcommand{\U}[1]{\underline{#1}}
            \newcommand{\X}{\!\mvec{X} (\cdot)}
            \newcommand{\cd}{(\cdot)}
            \newcommand{\Q}{\mbox{\bf Q}}
            \newcommand{\p}{\partial_t}
            \newcommand{\z}{\!\mvec{z}}
            \newcommand{\bu}{\!\mvec{u}}
            \newcommand{\rr}{\!\mvec{r}}
            \newcommand{\w}{\!\mvec{w}}
            \newcommand{\g}{\!\mvec{g}}
            \newcommand{\D}{I\!\!D}
            \newcommand{\se}[1]{_{\mvec{;}#1}}
            \newcommand{\sek}[1]{_{\mvec{;}#1]}}            
            \newcommand{\seb}[1]{_{\mvec{;}#1)}}            
            \newcommand{\ko}[1]{_{\mvec{,}#1}}
            \newcommand{\ab}[1]{_{\mvec{|}#1}}
            \newcommand{\abb}[1]{_{\mvec{||}#1}}
            \newcommand{\td}{{^{\bullet}}}
            \newcommand{\eq}{{_{eq}}}
            \newcommand{\eqo}{{^{eq}}}
            \newcommand{\f}{\varphi}
            \newcommand{\dm}{\diamond}
            \newcommand{\seq}{\stackrel{_\bullet}{=}}
            \newcommand{\st}[2]{\stackrel{_#1}{#2}}
            \newcommand{\om}{\Omega}
            \newcommand{\emp}{\emptyset}
            \newcommand{\bt}{\bowtie}
            \newcommand{\btu}{\boxdot}
            \newcommand{\bti}{\boxtimes}
            \newcommand{\btp}{\boxplus}
            \newcommand{\kl}{\mvec{\{}}
            \newcommand{\kr}{\mvec{\}}}
\newcommand{\Section}[1]{\section{\mbox{}\hspace{-.6cm}.\hspace{.4cm}#1}}
\newcommand{\Subsection}[1]{\subsection{\mbox{}\hspace{-.6cm}.\hspace{.4cm}
\em #1}}

\newcommand{\const}{\textit{const.}}
\newcommand{\vect}[1]{\underline{\ensuremath{#1}}}  
\newcommand{\abl}[2]{\ensuremath{\frac{\partial #1}{\partial #2}}}

\abstract\noindent
The balance equations of energy-momentum and spin together with Einstein's
field equations are investigated by the Liu procedure to find constraints
for the constitutive equations in such a way that the Second Law is satisfied. 
Special cases such as spinless systems and curvature insensitive materials
are shortly discussed.

\section{Introduction}

The balance equations of continuum physics are formulated for arbitrary
materials, that means, they present an underdetermined system of differential
equations which need for its solution additional equations, the material
equations which made the system of the balance equations solvable for a special
material. There are two different procedures: The material equations are
presupposed\footnote{The practitioner knows his material and chooses the
correct ansatz.} and solved together
with the balances asking later, if the Second Law is satisfied for all
positions
and times. The second method takes the Second Law into consideration before
solving the balances\footnote{The theorist looks for a class of materials in
agreement with the Second Law.}
asking for material equations supplementing the balances equations so that
their common solution is in agreement with the Second Law. 
\vspace{.3cm}\newline
Here, we deal with the second method which is well known in non-relativistic
material theory \C{LL,12y}, but is up to now not strictly
developed in general-relativistic constitutive theory, except of an attempt
\C{HMR}. A reason for that delay
may be caused by the question, how to handle Einstein's field equations in this
framework. Precondition for taking Einstein's field equations into
consideration is that those and the balance equations are formulated in a
compatible form\footnote{The energy-momentum tensors are different the in field
and balance equations, that needs a modified Belinfante-Rosenfeld procedure,
here also taking the field equations into consideration \C{MP}.}.
Having done this, the question arises, if Einstein's field
equations fit in the well known procedure of generating the material class
being compatible with the Second Law\footnote{fot the non-relativistic case see
\C{MMD}}. This is the case due to the fact
that
Einstein's gravitational equations are differential equations of second order
for the metric. Thus, our approach is completely in accordance with Maugin's
opinion \C{MAUG}\footnote{\label{foot4}Maugin: ``First, from the formal point
of view, we may
consider Einstein's theory of general relativity as a generalized continuum
theory of the second gradient of the space-time metric.... Second, the Maxwell
stress and other such electromagnetic stress tensors....are none other than
purely spatial parts of the space-time energy-momentum tensor....Finally,
while we can have some doubt with special relativity, it is clear that general
relativity is a continuum theory from the start. This is more than enough to
ponder the formulation of continuum thermodynamics in its framework."}.
Encouraged by his remarks, we derive the Liu equations and the dissipation
inequality for the combined system of energy-momentum and spin balances --that
are the Mathisson-Papapetrou equations-- and Einstein's field equations.
Special cases are considered: spinless systems and curvature insensitive
materials.
\vspace{.3cm}\newline
The paper is organized as follows: After some introductory considerations on
tetrads and connexions, the balances and the gravitational equations are
written
down in (3+1)-decomposition. State spaces of first and second order are
introduced. The Liu equations and the dissipation inequality are derived for
a first order state space material. A discussion finishes the paper.

\section{Tetrads and Connexions}

Measuring values produced by measuring devices need for their description a 
frame of reference which is locally spanned by tetrads
$\{e^A_i\}$, $A,i = 1,...4$,  
consisting of three space-like vectors and one time-like vector numbered by
$A$ \footnote{The tetrad representation is well known. For more details we
refer to \C{1,2}.}.
These vectors represent a standard clock and
three standard measuring rods\footnote{That is the reason why we introduce
tetrads.}. The field of tetrads $\{e^A_i(x^a)\}$ is called an {\em observer
field} or a {\em system of reference}.
\vspace{.3cm}\newline
The corresponding contravariant components of the tetrad $e^A_i$,
denoted by $e^k_A$, are defined by
\bee{F4}
e^k_Ae^A_i\ =\ \delta^k_i,\quad\longleftrightarrow\quad
e^k_Ae^B_k\ =\ \delta^B_A.
\ee
The tetrad $e^k_A$ is presupposed to be {\em ortho-normalized} with respect to
the metric $g_{ik}$ of a pseudo-Riemannian space
\bee{F3}
g_{ik}(x^a)e^i_A(x^a)e^k_B(x^a) \ =\ \eta_{AB},\qquad 
(\eta_{AB})\ :=\
\left (
\begin{array}[c]{cc}
-{\bf 1} & \mvec{0}\\
\mvec{0}^{\top}& 1 
\end{array}
\right ).
\ee
The matrix $(\eta_{AB})$ has the shape of the Minkowski metric.
Multiplying \R{F3}$_1$ by $e^A_i e^B_k$ and using \R{F4} results in
\bee{F4a}
g_{ik}\ =\ \eta_{AB}e^A_ie^B_k 
\quad\longleftrightarrow\quad
g^{mn}\ =\ \eta^{CD}e^m_Ce^n_D.
\ee
The 16 components $\{e_A^i\}$ or $\{e_i^A\}$, are 
restricted by 10 constraints, \R{F3}$_1$ or \R{F4a}. Thus for a given metric,
10 components of the system of reference are fixed. 
The tetrads depend on the event, whereas $(\eta_{AB})$ does not.
\vspace{.3cm}\newline
The tetrad components $e_A^i$ and $e_i^A$, respectively, are matrices 
transforming tensor components at each point of the pseudo-Riemannian space.
For a tensor of first order, we have the following
\bee{F5}
a_B\ :=\ e^j_Ba_j \quad\longleftrightarrow\quad
a_j\ =\ e^B_ja_B,
\ee
and generally, we obtain for tensors of higher order
\bee{F8} 
T^{A...}\ _{B...}\ =\ T^{c...}\ _{d...}e^A_c e^d_B...,\qquad
T^{a...}\ _{b...}\ =\ T^{C...}\ _{D...}e^a_C e^D_b...\ .
\vspace{.3cm}\ee
We now consider local transformations $L^B_A(x^a)$ of the tetrad 
components
\bee{F8a}
\mbox{at }x^a :\hspace{1cm}
\st{*}{e}{^B_j}\ =\ L^B_A(*\dm)\st{\dm}{e}{^A_j},
\qquad
L_B^C(\dm *)L_A^B(*\dm)\ =\ \delta^C_A .
\ee
Inserting \R{F8a} into \R{F5}, results in
\bee{F8b}
\st{\dm}{a}_ A\ =\ L^B_A(*\dm) \st{*}{a}{_B}.
\vspace{.3cm}\ee
The properties of the matrix $L^B_A$ are connected to the ortho-normalized 
tetrads: Inserting \R{F8a} into \R{F4a} results in
\bee{F8d} 
e^A_i\ \equiv\ \st{*}{e}^A_i,\qquad
g_{ik}\ =\ \eta_{AB} L^A_C \st{\dm}{e}{^C_i} L^B_D\st{\dm}{e}{^D_k}.
\ee
The Principle of Relativity now asserts that all systems of reference which are
compatible with the given metric $g_{ik}$ are equivalent to each other, i.e.
that $\st{\dm}{e}{^A_j}$ in \R{F8a} belongs also to an ortho-normalized 
tetrad as $e^B_j$ in \R{F4a} does. According to \R{F8d}, this demand is
achieved by the setting
\bee{F8e}
\eta_{AB} L^A_C(x^a) L^B_D(x^a)\ \doteq\ \eta_{CD}.
\ee
This equation is the definition for $L^A_B(x^a)$ to be a Lorentz 
transformation at each point $(x^a)$ . Thus, $L^A_B(x^a)$ belongs to the 
group of local space-time dependent Lorentz transformations and transforms an
orthogonal-normalized tetrad into an orthogonal-normalized one at each $x^a$.  
\vspace{.3cm}\newline
In this paper, we consider the (anholonomic) tetrad representation of General
Relativity, instead of the (holonomic) metric one. Accordingly, one
has to work in a framework, in which the {\em Principle of General Relativity}
is realized by the covariance of the basic laws with respect to local Lorentz
transformations, instead of general coordinate transformations (what requires
the definition of the Lorentz connexion)\footnote{To avoid
misunderstandings, it should be mentioned that the tetrads here
introduced as anholonomic coordinates differ as well from those introced in
tetrad theories of gravitation, where they are considered as gravitational
potentials, as from those ones introduced as so-called directors for
describing the spin of the relativistic continua by Maugin and Eringen
\C{MAEH}.}. This is  the relativistic version of
the {\em Principle of Material-Frame Indifference}
(or objectivity)\footnote[9]{For the
special-relativistic theory, being covariant under rigid Lorentz
transformations, this was already shown by Maugin and Eringen \C{MAEH1}.
Later on in a series of papers in JMP \C{MAU}, Maugin considered the 
space-time covariant description of constitutive equations also in general
relativity as the relativistic replica of the classical notion of objectivity  
where covariance means: covariance with respect to general coordinate or
local Lorentz transformations. For the non-relativistic case see
\C{8,SB,BS,11}.}.
\vspace{.3cm}\newline
The covariant derivative $a^A_B{\ab{m}}$ is related to the partial one
$a^A_B{\ko{m}}$ by the connexion $\om^A_{mP}$ as follows\footnote[10]{for more
details concerning the tetrad calculus see here and in the sequel \C{2}}
\bee{F17}    
a^A_B{\ab{m}}\ =\ a^A_B{\ko{m}} + \om^A_{mP}a^P_B - \om^Q_{mB}a^A_Q,
\ee
and the connexion $\om^Q_{mB}$ is defined by
\bee{F22}
0\ =\ e^A_i{\ko{m}} - \Gamma^q_{mi}e_q^A +\om^A_{mQ}e^Q_i. 
\ee
Here $\Gamma^q_{mk}$ is the connexion belonging to the metric $g_{ik}$ in
\R{F4a}$_1$ \C{3}
\bee{F22a} 
\Gamma^b_{mi}\ =\ \frac{1}{2}(g_{mq,i} + g_{iq,m} - g_{mi,q})g^{bq}.
\ee
According to \R{F4a}$_1$, \R{F22} and \R{F22a}, the connexion is a function of
the tetrads and their first partial derivatives
\bee{F22b}
\om^A_{mQ}\ =\ \omega^A_{mQ}(e^D_j , e^D_{j,k}).
\ee

\section{Balance Equations\label{BE}}

In this section, we start out with the well-known general-relativistic 
balance and field equations of continuum physics. These balance equations
are gene\-rated from the special-relativistic balances by applying 
Einstein's {\em Principle of Equivalence},
mathematically realized by
the Princple of Minimal Coupling which asserts that the partial derivatives
in the special-relativistic balances have to be replaced by covariant ones
with respect to local Lorentz transformations \C{MK}.

\subsection{The general case}

Thus, we obtain in a pseudo-Riemannian space the following thermodynamical
balances:
\newline{\bf Particle number density:}
\bee{1y}
N^A{\ab{A}}\ =\ 0,
\quad N^A\ :=\ \frac{1}{c^2}nu^A .
\ee
Here, $u^A$ is the material 4-velocity of the particles and $c$ the vacuum
velocity of light.\newline
{\bf Energy-momentum:} 
\bee{2y}
T^{AB}{\ab{B}}\ =\ f^A .
\ee
Here, $T^{AB}$ is the energy-momentum tensor and $f^A$ the external
4-force density.\newline 
{\bf Spin:}
\bee{2ya}
S_{BC}^{A}{\ab{A}}\ =\ m_{BC}\ =\ -m_{CB}.
\ee
Here, $S_{BC}^{A}$ is the spin tensor and $m_{BC}$ the external 4-momentum
density.\newline
{\bf Entropy:}
\bee{3y}
S^A{\ab{A}} - \varphi\ =\ \sigma \geq 0
\ee
with the 4-entropy vector $S^A$, the entropy supply $\varphi$ and the
non-negative entropy production $\sigma$ \footnote[11]{All these quantities are
densities.}. This inequality represents the 
Second Law of classical field theories, called the {\em dissipation
inequality} \C{5a,14}.
\vspace{.3cm}\newline
Here, the constitutive theory is especially developed in the framework of
GRT\footnote[12]{General Relativity Theory}.
Consequently, we have to take into consideration Einstein's\newline 
{\bf Field Equations:}\C{3}
\bee{4y}
R_{AB} - \frac{1}{2}\eta_{AB} R =  \kappa\Theta_{AB}.
\ee
Here, $R_{AB}$ is the symmetric Ricci tensor and $R$ the curvature scalar
\bee{5y}
R_{AB} := R^C_{ACB},\qquad R := R^A_A ,
\ee
and $R^C_{ADB}$ is the curvature tensor of a pseudo-Riemannian space
\bey\nonumber
R^C_{ADB} &=& R^b_{mni}e^C_b e^m_Ae^n_D e^i_B\ =
\\ \label{a25} 
&=& (\Gamma^b_{mi\mvec{,}n}-\Gamma^b_{mn\mvec{,}i} +
\Gamma^j_{mi}\Gamma^b_{nj} - \Gamma^j_{mn}\Gamma^b_{ij})\ 
e^C_b e^m_Ae^n_D e^i_B.
\eey
As in the appendix \ref{CTTR} derived, the curvature tensor is linear in
the second partial derivatives of the tetrads
\bee{a25ab}
R^b_{mni}\ =\ E^{buvw}_{mniG}(e^A_j)e^G_{u}{\ko{vw}} +
F^b_{mni}(e^A_j,e^A_{j,p}),
\ee
and depends also on their first derivatives and on the tetrads themselves.  
Completely written in tetrad components, we have 
\bee{a25z}
R^C_{ADB}\ =\ G^{Cuvw}_{ADBG}(e^A_j,e^A_{j,p})e^G_{u}{\ko{vw}} +
H^C_{ADB}(e^A_j,e^A_{j,p}).
\vspace{.3cm}\ee
In accordance with \R{4y}, the {\em gravitation generating energy-momentum
tensor} $\Theta_{AB}$ has to be symmetric and divergence-free
\bee{7y}
\Theta_{AB} = \Theta_{BA}\ \wedge\ \Theta^{AB}{\ab{B}} = 0 .
\ee
The field equations \R{4y} yield by multiplication with $\eta^{CA}$
and by use of \R{5y}$_2$
\bee{13d1y}
R = -\kappa\Theta^B_B \quad\longrightarrow\quad
R_{AB} = \kappa\Theta_{AB} - \frac{1}{2}\eta_{AB} \kappa\Theta^C_C.
\vspace{.3cm}\ee
Obvious is, that the gravitation generating energy-momentum tensor
$\Theta_{AB}$ is different from that in the energy-momentum balance,
because $T_{AB}$ does not satisfy \R{7y}. Therefore the question
arises: what is the relation between $\Theta_{AB}$ and $T_{AB}$?\footnote[13]{For
the Special Relativity Theory, this question is answered in \C{BELROS}.}
An additional question is: are the balance equations \R{2y} and
\R{2ya} compatible with the field equations \R{4y}? To answer
this question, we refer to \C{MP}, where we proved the following
\vspace{.3cm}\newline
$\blacksquare$ {\sf Proposition:} If we presuppose that in GRT energy-momentum
and spin sa\-tis\-fy balance equations --\R{2y} and \R{2ya}-- then the
gravitation generating energy-momentum tensor in Einstein's field
equations is 
\byy{x7y} 
\Theta^{AB} &=& T^{AB} - \frac{1}{2}\Sigma^{CAB}{\ab{C}},
\\ \label{w7y}  
\Sigma^{CAB} &:=& S^{CAB} + S^{ABC} + S^{BAC},
\eey
and the external force density and the momentum density are
\byy{z7y} 
f^A &=& -\frac{1}{2}R^A_{CDE}S^{CDE},
\\ \label{y7y}
m^{BA}&=& 2T^{[BA]}.
\eey
Consequently, the corresponding balance equations of energy-momentum
and spin are the Mathisson-Papapetrou equations \C{MP1,MP2}
\byy{v7y} 
T^{AB}{\ab{A}} &=& -\frac{1}{2}R^B_{CDE}S^{CDE},
\\ \label{u7y}
S^{CAB}{\ab{C}} &=& 2T^{[AB]}.\hspace{5cm}\blacksquare\hspace{-3cm}
\eey
The gravitational equations \R{13d1y} of General Relativistic
Continuum Theory (GRCT) become
\bey\nonumber
R^{AB} + \frac{1}{2}\eta^{AB}\kappa\Theta^D_D &=&  \kappa\Big(T^{AB} - 
\frac{1}{2}( S^{CAB} + S^{ABC} + S^{BAC}){\ab{C}}\Big)=
\\ \label{t7y}
&=& \kappa\Big(T^{(AB)} - 
\frac{1}{2}(S^{ABC} + S^{BAC}){\ab{C}}\Big).
\eey
The trace of the gravitation generating energy-momentum tensor is
according to \R{t7y} and \R{13d1y}
\bee{s7y}
\Theta^D_D\ =\ T^D_D - S^{DC}_D{\ab{C}}.
\ee
Consequently, the Ricci tensor in \R{t7y} becomes
\bee{r7y}
R^{AB}\ =\ \kappa\Big(T^{(AB)} - 
\frac{1}{2}(S^{ABC} + S^{BAC}){\ab{C}} - \frac{1}{2}\eta^{AB}
\Big[T^D_D - S^{DC}_D{\ab{C}}\Big]\Big)
\ee
which is determined by the symmetric part of the energy-momentum
tensor and derivatives of the spin tensor.
The field equations \R{t7y}
become
\bee{q7y} 
(S^{ABC} + S^{BAC} + \eta^{AB}S^{DC}_D){\ab{C}}+\frac{2}{\kappa}R^{AB}
\ =\ 2 T^{(AB)} - \eta^{AB}T^D_D .
\ee
According to \R{q7y}, all derivatives appearing in the field equations are
on the lhs. This form of the field equations
is essential for the exploitation of the dissipation inequality
\R{3y} by the Liu procedure which is discussed in sect.\ref{DI}.

\subsection{(3+1)-Decomposition}

For the exploitation of the dissipation inequality, we need \R{v7y}, \R{u7y}
and \R{q7y} in (3+1)-decomposition. Starting out with the particle flux, the
energy-momentum tensor, the spin tensor and the 4-entropy \C{6,GR,HH},
we obtain
\byy{12yc}
N^A &=& \frac{1}{c^2}nu^A,
\\ \label{a12yc}
T^{AB} &=& \Big(\frac{1}{c^4}eu^B + \frac{1}{c^2}p^B\Big)u^A +
\frac{1}{c^2}u^Bq^A + t^{AB},
\\ \label{12yd}
S^{CAB} &=& \Big(\frac{1}{c^2}s^{AB} + \frac{1}{c^4}u^{[A}\Xi^{B]}\Big)u^C +
\frac{1}{c^2}u^{[A}\Xi^{B]C} + s^{CAB},
\\ \label{12ye}
S^A &=& \frac{1}{c^2}su^A + s^A.
\eey
In the sequel, we split the fields of the (3+1)-decomposition into two classes.
Using the projector $h^{AB}$ onto the sub-space perpendicular to the 
4-velocity $u^A$
\bee{13y}
h^{AB} := \eta^{AB} - \frac{1}{c^2}u^A u^B  = h^{BA},
\ee
the first class is defined by
\byy{8ya}
n &:=& N^A u_A,
\\ \label{9y}
e &:=& T_{AB}u^A u^B ,\qquad \varepsilon := e/c^2,
\\ \label{9ya}
s_{AB} &:=& S_{EF}^{C}h^E_A h^F_B u_C ,
\\ \label{9yb}
\Xi_A &:=& 2S_{EF}^{C}u_C u^E h^F_A.
\eey
Here $n$ is the {\em particle density}, $e$ the {\em energy density}, $s_{AB}$
the {\em spin density} and finally $\Xi_A$ the {\em spin density vector}.
\vspace{.3cm}\newline
The second class of fields of the (3+1)-decomposition is defined by \C{6,HH}
\byy{12y}
t^{AB} &:=& h^{AE}T_{EF}h^{FB},
\\ \label{a12y}
p^A &:=& h^{AE}T_{FE}u^F,
\\ \label{12y1} 
q^A &:=& h^{AE}T_{EF}u^F,
\\ \label{12ya}
s_{AB}^{C} &:=& S_{EF}^{G}h^E_A h^F_B h^C_G,
\\ \label{12yb}
\Xi_B^{A} &:=& 2S_{EF}^{G}h_G^A u^E h^F_B,
\\ \label{a12yb}
s &:=& S^Au_a,\qquad s^A\ :=\ S^B h^A_B. 
\eey
Here $t^{AB}$ is the {\em stress tensor}, $p^A$ the {\em momentum flux
density}, $q^A$ the {\em energy flux density},
$s_{AB}^{C}$ the {\em couple stress}, $\Xi_B^{A}$ the {\em spin stress},
$s$ the {\em entropy density} and $s^A$ the {\em entropy flux}.
\vspace{.3cm}\newline
We now write down the balance equations in (3+1)-decomposition.
According to \R{v7y} and \R{a12yc}, we obtain the energy-momentum balance
\bey\nonumber
\frac{1}{c^2}\Big(p^B\ab{A}u^A + u^Bq^A\ab{A}\Big) + t^{AB}\ab{A} +
\frac{1}{c^2}\Big(p^Bu^A\ab{A} + u^B\ab{A}q^A\Big) +
\frac{1}{c^4}(eu^Bu^A)\ab{A}\ =
\\ \label{D1}
=\ -\frac{1}{2}R^B_{CDE}S^{CDE}.
\eey
The spin balance results from \R{u7y} and \R{12yd}
\bee{D2}
\frac{1}{c^2}u^{[A}\Xi^{B]C}\ab{C} + s^{CAB}\ab{C} +
\frac{1}{c^2}u^{[A}\ab{C}\Xi^{B]C} +
\Big[\Big(\frac{1}{c^2}s^{AB} + \frac{1}{c^4}u^{[A}\Xi^{B]}\Big)u^C\Big]\ab{C}\
=\ 2T^{[AB]}, 
\ee
and from \R{3y} and \R{12ye} follows the dissipation inequality
\bee{D3}
\frac{1}{c^2}s\ab{A}u^A + s^A\ab{A} + \frac{1}{c^2}su^A\ab{A}\ \geq\ \varphi.
\ee
The field equations \R{q7y} by use of \R{12yd} result in a long-winded
expression
\bey\nonumber
\frac{1}{c^2}u^{[B}\Xi^{C]A}\ab{C} + s^{ABC}\ab{C} +
\frac{1}{c^2}u^{[B}\ab{C}\Xi^{C]A} +
\Big[\Big(\frac{1}{c^2}s^{BC} + \frac{1}{c^4}u^{[B}\Xi^{C]}\Big)u^A\Big]\ab{C}+
\\ \nonumber
+ \frac{1}{c^2}u^{[A}\Xi^{C]B}\ab{C} + s^{BAC}\ab{C} +
\frac{1}{c^2}u^{[A}\ab{C}\Xi^{C]B} +
\Big[\Big(\frac{1}{c^2}s^{AC} + \frac{1}{c^4}u^{[A}\Xi^{C]}\Big)u^B\Big]\ab{C}+
\\ \nonumber
+ \eta^{AB}\Big\{
\frac{1}{c^2}u_{[D}\Xi^{C]D}\ab{C}\! + s^{DC}_D{\ab{C}}\! +
\frac{1}{c^2}u_{[D}{\ab{C}}\Xi^{C]D}\! +
\Big[\Big(\frac{1}{c^2}s^{C}_D + \frac{1}{c^4}u_{[D}\Xi^{C]}\Big)u^D\Big]\ab{C}
\Big\}\! \! +
\\ \label{D4}
+ \frac{2}{\kappa}R^{AB}\ =\ 2 T^{(AB)}- \eta^{AB}T^D_D.\hspace{.5cm}
\vspace{.3cm}\eey
The system of diffential equations \R{D1} to \R{D4} is valid for arbitrary
materials. For solving it, we need constitutive equations and the Second Law
\R{D3} has to be taken into account.

\section{State Spaces}

\subsection{Basic and constitutive fields}

The balances \R{1y}, \R{2y} and \R{2ya} and the field equations \R{4y} 
represent equations to calculate the 58 fields $N^A , T^{AB}, S^A_{BC},
S^A$ and $e^A_i$ by taking the constraint \R{F4a}
into consideration\footnote[14]{4+16+24+4+10=58. The 10 $g_{ik}$ are
replaced by 10 bilinear combinations of the 16 $e^A_i$ according to \R{F4a}.
}. 
Additionally, we have a further constraint by the dis\-sipation inequa\-li\-ty
\R{3y}. Balances and field equations are valid for arbitrary materials,
e.g. non-Newtonian fluids, spin-fluids, solids, liquid crystals.
To solve the system of differential equations combined of the balances and the 
field equations, we have to introduce constitutive equations. Therefore, we 
have to split these
58 fields into those which represent the independent variables,
called the {\em basic fields}\footnote[15]{often named the
{\em wanted fields}},
and those which are material dependent ones, named the {\em constitutive 
fields} \C{7}.
\vspace{.3cm}\newline
Like in non-relativistic continuum physics, we choose as basic fields
the first class \R{8ya} to \R{9yb} and additional the field of tetrads which
according to \R{F8a}$_1$ are tensors of first order in the tetrad indices
\bee{8y}
\mvec{z} = (n, u^A , e , s_{AB}, \Xi_A, e^A_i)\ =:\ (\mvec{w},e^A_i).
\ee
These basic variables do not contain any derivative.
Taking the normalization of the 4-velocity
\bee{10y}
u^A u_A = c^2 \quad\longrightarrow\quad u^A u_{A|B}\ =\ 0 
\ee
and \R{F4a} into account, the number of the basic fields in \R{8y} 
is 21.\footnote[16]{1+3+1+3+3+10=21}
\vspace{.3cm}\newline
According to the splitting into basic and constitutive fields, we now
have to introduce 58 - 21 = 37 constitutive fields\footnote[17]{9+3+3+9+9+4=37}. 
These are
\bee{11y}
\mvec{M} = (t^{AB}, p^A, q^A , s^C_{AB}, \Xi^A_B, S^A).
\ee
Basic and constitutive fields are tensors of different order, that means, they 
are covariantly defined.

\subsection{Constitutive mappings\label{CM}}

For describing constitutive properties, we have to introduce 
{\em material mappings}\footnote[18]{also called constitutive mappings \C{7}} 
which connect the constitutive fields \R{11y}, $\mvec{M}$, with the basic
fields \R{8y}, $\mvec{z}$, thus characterizing the material.
Obvious is, that the basic fields as the independent variables do not describe 
materials sufficiently in non-equilibrium.
Consequently, we have to extend $\mvec{z}$ by additional variables 
$\mvec{\zeta}$ consisting of derivatives of the basic fields\footnote[19]{An
other
possibility of extending the basic variables is to take instead of derivatives
constitutive fields into the extension, e.g. $t^{AB}$ and $q^A$, see: Extended
Thermodynamics \C{EXT1,EXT2}.}
\bee{M1}
\mvec{\zeta}\ =\ (\mvec{z}_{|m}, \mvec{z}_{|m|n},... ).
\ee
The order of the additional derivatives which are enclosed in $\mvec{\zeta}$
depends on the considered material. Consequently, the {\em set of constitutive
properties} \R{11y} depends on $(\mvec{z},\mvec{\zeta})$. which is called the 
covariantly defined
{\em state space}\footnote[20]{or constitutive space} and which spans the
domain
of the material mapping  ${\cal M}$
\bee{G5}
\mvec{M}\ =\ {\cal M}(\mvec{z},\mvec{\zeta}).
\ee
Its range $\mvec{M}$ is spanned by the constitutive properties \R{11y}. 
The entire equation \R{G5} is kown as the {\em constitutive} or {\em material
equation} 
\bee{M3}
\mvec{M}\ =\ {\cal M}(\mvec{z},\mvec{z}_{|m}, \mvec{z}_{|m|n},...).
\vspace{.3cm}\ee
If we use partial derivatives instead of covariant ones, the connexion has to
be added to the additional variables
\bee{M2}
\mvec{\zeta}'\ = \ (\mvec{z}{\ko{m}}, \mvec{z}{\ko{mn}},...,\om^B_{mA},
\om^B_{mA}{\ko{n}},...),
\ee
and the constitutive equation becomes \C{10}
\bee{M3a}
\mvec{M}\ =\ {\cal M}(\mvec{z},\mvec{z}{\ko{m}}, \mvec{z}{\ko{mn}},...,
\om^B_{mA},\om^B_{mA}{\ko{n}},... ).
\ee
According to \R{F22b}, the connexion itself depends on the tetrads and their
first partial derivatives which are included in the 
$(\mvec{z}_{,m}, \mvec{z}_{,m,n},...)$. Consequently, $\om^B_{mA}$ can be
cancelled in the set of th state space variables
\bee{M4}
\mvec{M}\ =\ {\cal M}(\mvec{z},\mvec{z}_{,m}, \mvec{z}_{,m,n},... ).
\ee
In this form of the constitutive equation, the state space does not consist
of tensors in contrast to \R{M3}, but it contains all non-tensorial quantities 
so that by combination of them, the tensorial constitutive equation \R{M3} 
can be rediscovered. Here, we prefer the non-tensorial representation of the
constitutive equation, because curvature and Ricci tensor are functions of
state space variables which according to \R{a25z} are partial derivatives.

\subsection{First and second derivatives state spaces}

Taking \R{F22}$_2$ into account,
the additional variables \R{M1} of first derivative state spaces are
\bee{12sa}
\mvec{\zeta}'^I\ \equiv\ (\mvec{z}\ko{m})
=\ (n\ko{m}, u_{A}{\ko{m}}, \varepsilon\ko{m},
s_{AB}{\ko{m}} ,\Xi_{A}{\ko{m}}, e^A_i{\ko{m}})\ =\ 
(\mvec{w}{\ko{m}},e^A_i{\ko{m}}).
\ee
Consequently, the state space of first partial derivatives becomes
\bey\nonumber
{\cal Z}^{I} &=&
(\mvec{z},\mvec{z}\ko{m})\ = (\mvec{w},\mvec{w}\ko{m},e^A_i,e^A_i{\ko{m}})\ =\
\\ \label{12s1}
&=&(n, u^A , e , s_{AB}, \Xi_A, e^A_i,n\ko{m}, u_{A}{\ko{m}}, 
\varepsilon\ko{m},s_{AB}{\ko{m}} ,\Xi_{A}{\ko{m}}, e^A_i{\ko{m}}).
\eey
The state space ${\cal Z}^I$ is a local object defined at $x^a$, spanned by
independent variables --the wanted basic fields and their first derivatives--
generating locally the constitutive fields \R{11y} by the material mapping
\R{G5}.  
\vspace{.3cm}\newline
The additional variables of second order are
\bey\nonumber
\mvec{\zeta}'^{II} &\equiv& (\mvec{z}\ko{m},\mvec{z}_{\ko{mn}})\ =\
(\mvec{w}\ko{m},\mvec{w}_{\ko{mn}},e^A_i{\ko{m}},e^A_i{\ko{mn}})\ =
\\ \label{12s2}
&=& ( \mvec{\zeta}'^{I}, n_{\ko{mn}}, u_A{\ko{mn}},\varepsilon{\ko{mn}},
s_{AB}{\ko{mn}} ,\Xi_A{\ko{mn}} ,e^A_i{\ko{mn}} ),
\eey
resulting in the state space of second order
\bey\nonumber
{\cal Z}^{II}\ =\ \Big(n, u^k , e , s_{AB}, \Xi_A, e^A_i,
n\ko{m}, u_{A}{\ko{m}}, 
\varepsilon\ko{m},s_{AB}{\ko{m}} ,\Xi_{A}{\ko{m}}, e^A_i{\ko{m}},
\\ \label{12s3}
n_{\ko{mn}}, u_A{\ko{mn}},\varepsilon{\ko{mn}},
s_{AB}{\ko{mn}} ,\Xi_A{\ko{mn}} ,e^A_i{\ko{mn}}\Big).
\vspace{.3cm}\eey
Having introduced the concepts of state space and material mapping, we are able
to consider the system of differential equations \R{D1} to \R{D4} on the chosen
state spaces.

\subsection{Balances and field equations on the state space\label{FESS}}

We especially consider a state space of first order \R{12s1} and write down
the system \R{D1} to \R{D4} in which expressions like $\boxdot^{CB}{\ab{A}}$
appear. These expressions have according to \R{F5} and \R{F17} the form
\bee{D4a}
\boxdot^{CB}{\ab{A}}\ =\ \boxdot^{CB}{\ab{j}}e^j_A\ =\
\Big(\boxdot^{CB}\ko{j} + \om^C_{jQ}\boxdot^{QB} + \om^B_{jQ}\boxdot^{CQ}\Big)
e^j_A.
\ee
Taking \R{F22b} into account, we obtain
\byy{D4b}
\boxdot^{CB}{\ab{A}}\ =\ \boxdot^{CB}\ko{j}e^j_A + 
\kl \boxdot^{CB}_A\kr (\mvec{z},\mvec{z}_{|m}),
\\ \label{D4c}
\kl \boxdot^{CB}_A\kr (\mvec{z},\mvec{z}_{|m})\ :=\
\Big(\om^C_{jQ}\boxdot^{QB} +\om^B_{jQ}\boxdot^{CQ}\Big)
e^j_A.
\eey
Because of \R{F22b} and the chosen state space of first order,
the bracket symbol
$\kl \boxdot^{CB}_A\kr$ does not depend on higher derivatives.
The partial derivative has the
following meaning, distinguishing between basic and additional variables
according to \R{12s1}
\bee{D5}
{\boxdot}^{CB}{\ko{j}}\ =\
\Big(
\frac{\partial{\boxdot^{CB}}}{\partial {w}{^k}}{w}{^k}{\ko{j}} +
\frac{\partial{\boxdot^{CB}}}{\partial {w}{^k}{\ko{m}}}
{w}{^k}{\ko{mj}} +
\frac{\partial{\boxdot^{CB}}}{\partial e^Q_k}e^Q_k{\ko{j}} +
\frac{\partial{\boxdot^{CB}}}{\partial e^Q_k{\ko{m}}}
e^Q_k{\ko{mj}}
\Big).
\vspace{.3cm}\ee
By use of \R{D5}, \R{D4b} and \R{a25z}, \R{D1} results in the
{\em energy-momentum balance on the state space}
\bey\nonumber
&&\hspace{.5cm}
\Big(\frac{1}{c^2}(\frac{\partial p^B}{\partial w^k{\ko{m}}}u^A
+\frac{\partial q^A}{\partial w^k{\ko{m}}}u^B) +
\frac{\partial t^{AB}}{\partial w^k{\ko{m}}}\Big)e^j_A w^k{\ko{mj}} +
\\ \nonumber
&&\hspace{2cm}
+\ \Big(\frac{1}{c^2}(\frac{\partial p^B}{\partial e^Q_{k}{\ko{m}}}u^A
+\frac{\partial q^A}{\partial e^Q_{k}{\ko{m}} }u^B) +
\frac{\partial t^{AB}}{\partial e^Q_{k}{\ko{m}} }\Big)e^j_A e^Q_{k}{\ko{mj}}+
\\ \nonumber
&&\hspace{5cm}
+\ \frac{1}{2}S^{CDE}G^{Bkmj}_{CDEQ}(e^A_q,e^A_{q,p})e^Q_{k}{\ko{mj}}\ =
\\ \nonumber
&&
=\ -\Big(\frac{1}{c^2}(\frac{\partial p^B}{\partial w^k} u^A
+\frac{\partial q^A}{\partial w^k} u^B) + 
\frac{\partial t^{AB}}{\partial w^k}\Big)e^j_A w^k{\ko{j}}-
\\ \nonumber
&&\hspace{2cm}
-\ \Big(\frac{1}{c^2}(\frac{\partial p^B}{\partial e^Q_{k}}u^A
+\frac{\partial q^A}{\partial e^Q_{k}}u^B) +
\frac{\partial t^{AB}}{\partial e^Q_{k}}\Big)e^j_A e^Q_{k}{\ko{j}}\ -
\\ \nonumber
&&-\ \frac{1}{c^2}\Big(\kl p^B_A\kr u^A + u^B\kl q^A_A\kr\Big)
- \kl t^{AB}_A\kr\ - \frac{1}{2} S^{CDE}H^B_{CDE}(e^A_j,e^A_{j,p})\ -
\\ \label{D6}
&&\hspace{3.5cm}
- \frac{1}{c^2}\Big(p^Bu^A\ab{A} + u^B\ab{A}q^A\Big) +
\frac{1}{c^4}(eu^Bu^A)\ab{A}.
\eey
The lhs of the energy-momentum balance \R{D6} is linear in the higher
derivatives $w^k{\ko{mj}}$ and $e^Q_{k}{\ko{mj}}$, whereas no higher
derivatives appear on the rhs which depends on the state space variables of
the first order state space \R{12s1}. 
\vspace{.3cm}\newline
Starting out with \R{D3}, the dissipation inequality becomes with \R{D4b}
\bee{D6a}
\frac{1}{c^2}s{\ko{j}}e^j_A u^A + s^A{\ko{j}} e^j_A + \kl s^A_A\kr \geq
\varphi - \frac{1}{c^2}su^A\ab{A}.
\ee
By taking \R{D5} into account, we obtain the {\em the dissipation inequality
on the state space}
\bey\nonumber
\Big(\frac{1}{c^2}\frac{\partial{s}}{\partial {w}{^k}{\ko{m}}}u^A +
\frac{\partial{s^A}}{\partial {w}{^k}{\ko{m}}}\Big)e^j_A{w}{^k}{\ko{mj}} + 
\Big(\frac{1}{c^2} \frac{\partial{s}}{\partial e^Q_k{\ko{m}}}u^A +
\frac{\partial{s^A}}{\partial e^Q_k{\ko{m}}}\Big)e^j_Ae^Q_k{\ko{mj}}\ \geq
\\ \nonumber
\geq\ -\Big(\frac{1}{c^2}\frac{\partial{s}}{\partial w^k}u^A +
\frac{\partial{s^A}}{\partial w^k}\Big)e^j_A{w}{^k}{\ko{j}} -
\Big(\frac{1}{c^2} \frac{\partial{s}}{\partial e^Q_k}u^A +
\frac{\partial{s^A}}{\partial e^Q_k}\Big)e^j_Ae^Q_k{\ko{j}} +\hspace{.5cm}
\\ \label{D6b}
+\varphi -\kl s^A_A\kr -\frac{1}{c^2}su^A\ab{A}.\hspace{.8cm}
\vspace{.3cm}\eey
The pretty long-winded energy-momentum balance \R{D6} and the dissipation
inequality \R{D6b} can be written in a symbolic form 
\byy{D7}
\mbox{energy-momentum:}\hspace{1.5cm}
^{11}{\cal A}^{Bmj}_k w^k{\ko{mj}} +\ ^{12}{\cal A}^{Bkmj}_Q
e^Q_{k}{\ko{mj}} &=& ^1{\cal C}^B.
\\ \label{D10}
\mbox{dissipation inequality:}\hspace{2cm}
^1{\cal B}_k^{mj}w^k{\ko{mj}} + ^2{\cal B}_Q^{kmj} e^Q_{k}{\ko{mj}}
&\geq& {\cal D}.
\eey
The coefficients $^{11}{\cal A}$, $^{12}{\cal A}$, $^1{\cal C}$, $^1{\cal B}$,
$^2{\cal B}$ and ${\cal D}$ can be directly read off from \R{D6} and \R{D6b}.
\vspace{.3cm}\newline
The spin balance \R{D2} and the field equations \R{D4} can be treated in the
same
manner as the energy-momentum balance and the dissipation inequality. Their
symbolic shape is as follows
\byy{D8}
\mbox{spin:}\hspace{3cm}
^{21}{\cal A}^{ABmj}_k w^k{\ko{mj}} + ^{22}{\cal A}^{ABkmj}_Q e^Q_{k}{\ko{mj}}
&=& ^2{\cal C}^{AB},
\\ \label{D9}
\mbox{field equations:}\hspace{1cm}
^{31}{\cal A}^{ABmj}_k w^k{\ko{mj}} +\ ^{32}{\cal A}^{ABkmj}_Q e^Q_{k}{\ko{mj}}
&=& ^3{\cal C}^{AB}.
\eey
The definitions of $^{21}{\cal A}$,$^{22}{\cal A}$, $^{31}{\cal A}$,
$^{32}{\cal A}$, $^2{\cal C}^{AB}$ and $^3{\cal C}^{AB}$ can be obtained from
\R{D2} and \R{D4}
by the same treatment which was performed with the energy-momentum balance. 
We will not perform this procedure here, because our main aim is to demonstrate
that an exploitation of the dissipation inequality is also possible in General
Relativity Theory.
\vspace{.3cm}\newline
In the sequel we need the balances of
energy-momentum and spin, the gravitational field equations and the dissipation
inequality  --\R{D7} to \R{D9}-- in a common matrix formulation
\byy{D12}
\left(
\begin{array}{lr}
^{11}{\cal A}^{Bmj}_k & ^{12}{\cal A}^{Bkmj}_Q\\
^{21}{\cal A}^{ABmj}_k & ^{22}{\cal A}^{ABkmj}_Q  \\
^{31}{\cal A}^{ABmj}_k &^{32}{\cal A}^{ABkmj}_Q
\end{array}
\right)
\left(
\begin{array}{c}
w^k{\ko{mj}}\\
e^Q_{k}{\ko{mj}}
\end{array}
\right)
&=&
\left(
\begin{array}{c}
^1{\cal C}^B \\
^2{\cal C}^{AB} \\
^3{\cal C}^{AB} 
\end{array}
\right)
\\ \label{D13}
\left(
\begin{array}{lr}
^1{\cal B}_k^{mj} & ^2{\cal B}_Q^{kmj}
\end{array}
\right)
\left(
\begin{array}{c}
w^k{\ko{mj}}\\
e^Q_{k}{\ko{mj}}
\end{array}
\right)
&\geq&
{\cal D}.
\eey
The dissipation inequality \R{D13} enforces that the ${\cal A}$, ${\cal B}$,
${\cal C}$ and ${\cal D}$ are not arbitrary. The exploitation of the
dissipation inequality, the Liu procedure, is considered in the next section.

\section{Exploitation of the Dissipation Inequality\label{DI}}

The usual way to introduce the Second Law --the dissipation inequality \R{3y}--
into the exploitation of the balance equations --here \R{1y}, \R{v7y}, \R{u7y}
and \R{q7y}-- is to solve this system of coupled differential equations and 
afterwards to check, if the dissipation inequality \R{3y}
is satisfied for all events $x^a$. Apart from the complexity of this procedure,
it is not possible, because for solving the balances, we need
constitutive equations determining the constitutive fields \R{11y}.
But here, we are not interested in solving these balances, but we are looking
for general restrictions to the constitutive equations by the dissipation
inequality.

\subsection{The Liu procedure}

On account of the unknown constitutive equations, we
cannot solve the balance equations, but we can ask for the conditions which
must be satisfied by the constitutive equations, so that the dissipation 
inequality is locally valid for all events. 
The Liu procedure is such a tool for exploiting the dissipation inequality with
regard to the balance and field equations resulting in restrictions for the
constitutive equations \C{LL,12y}\footnote[21]{An other procedure to take the
Second Law into consideration is that of Coleman and Noll \C{COLNOL}.}.
The procedure removes the {\em higher derivatives}
which are outside of the state space, and it generates constraints
for the derivatives of the constitutive properties \R{11y} in the state space.
Here, we present the Liu procedure in a symbolical formalism.
\vspace{.3cm}\newline
The balances of energy-momentum \R{v7y} and of spin \R{u7y}, the field
equations \R{q7y} and the dissipation inequality \R{2ya} have the following 
shape
\bee{L1}
\bti^{AC}\ab{C}(\Delta^M)\ =\ \otimes^A,\qquad
\btp^C\ab{C}(\Delta^M)\ \geq\ \oplus. 
\ee 
Here, $\Delta^M$ is the symbol for the state space variables 
$(\mvec{z},\mvec{\zeta})$ in \R{G5}.
\vspace{.3cm}\newline
Starting out with \R{L1}$_1$, we obtain according to \R{D4b}
\bee{L2}
\bti^{AC}\ab{C}\ =\ \bti^{AC}\ko{j}e^j_C + \kl\bti^{AC}_C\kr \ =\
\otimes^A.
\ee
We now split the state space variables
\bee{L2a}
\Delta^M\ =\ (\Delta^M_-,\Delta^M_+)
\ee
into those ones $\Delta^M_+$ whose derivatives $\Delta^M_{+}{\ko{j}}$, the
higher derivatives, are out of the state space.
Then, we split \R{L2} into that part which is linear in the higher
derivatives, and into the remaining one which forms a part of the rhs
of the following equation 
\bee{L3}
e^j_C \frac{\partial\bti^{AC}(\Delta^M)}{\partial \Delta^M_+}
\Delta^M_{+}{\ko{j}}\
=\ -e^j_C \frac{\partial\bti^{AC}(\Delta^M)}{\partial \Delta^M_-}
\Delta^M_{-}{\ko{j}} - \kl\bti^{AC}_C\kr
+ \otimes^A\ =:\ \boxdot^A.
\ee
Analogously, the dissipation inequality \R{L1}$_2$ results in
\bee{L5}
e^j_C\frac{\partial \btp^C(\Delta^M)}{\partial \Delta^M_+}
\Delta^M_{+}{\ko{j}}\ \geq\ 
-e^j_C\frac{\partial \btp^C(\Delta^M)}{\partial \Delta^M_-}
\Delta^M_{-}{\ko{j}} - \kl\btp^{C}_C\kr
+ \oplus\ =:\ \odot .
\ee
Now we formulate
\vspace{.3cm}\newline
$\blacksquare$ {\sf Liu's Theorem}\C{LL,12y}: There are functions of the state
space variables $\Lambda_{A}(\Delta^M)$ generating the {\em Liu equations}
\bee{L6}
\Lambda_{A}e^j_C\frac{\partial\bti^{AC}}{\partial \Delta^M_+}\ =\ 
e^j_C\frac{\partial \btp^C}{\partial \Delta^M_+}\ \longrightarrow\
\Lambda_{A}\frac{\partial\bti^{AB}}{\partial \Delta^M_+}\ =\ 
\frac{\partial \btp^B}{\partial \Delta^M_+}
\ee
and the {\em reduced dissipation inequality}
\bee{L7}
\Lambda_{A}\boxdot^A \geq\ \odot ,
\ee
removing the higher derivatives $\Delta^M_{+}{\ko{j}}$.
\hfill$\blacksquare$
\vspace{.3cm}\newline
Liu equations and the reduced dissipation inequality represent constraints
for the constitutive quantities $\bti^{AC}$, $\btp^C$, $\otimes^A$ and
$\oplus$ which we derive in the next section.

\subsection{Liu equations and reduced dissipation inequality}

According to \R{L3} and \R{L5}, the lhs of the matrix equations \R{D12} and
\R{D13} are linear in the higher derivatives. Consequently, we can apply the
Liu procedure which according to \R{L6}$_2$ and \R{L7} results in the following
matrix equations
\byy{L8}
\left(
\begin{array}{lcr}
^1\Lambda_B & ^2\Lambda_{AB} & ^3\Lambda_{AB}
\end{array}
\right)
\left(
\begin{array}{lr}
^{11}{\cal A}^{Bmj}_k & ^{12}{\cal A}^{Bkmj}_Q\\
^{21}{\cal A}^{ABmj}_k & ^{22}{\cal A}^{ABkmj}_Q \\
^{31}{\cal A}^{ABmj}_k &^{32}{\cal A}^{ABkmj}_Q
\end{array}
\right)
\! &=&\! \! \!
\left(
\begin{array}{lr}
^1{\cal B}_k^{mj} & ^2{\cal B}_Q^{kmj}
\end{array}
\right),
\\ \label{L9}
\left(
\begin{array}{lcr}
^1\Lambda_B & ^2\Lambda_{AB} & ^3\Lambda_{AB}
\end{array}
\right)
\left(
\begin{array}{c}
^1{\cal C}^B \\
^2{\cal C}^{AB} \\
^3{\cal C}^{AB} 
\end{array}
\right)
&\geq&
{\cal D}.
\vspace{.3cm}\eey
Thus, we obtain the following Liu equations
\byy{L9a} 
^1{\cal B}_k^{mj} &=& ^1\Lambda_B\ ^{11}{\cal A}^{Bmj}_k +
^2\Lambda_{AB}\ ^{21}{\cal A}^{ABmj}_k +^3\Lambda_{AB}\ ^{31}{\cal A}^{ABmj}_k,
\\ \label{L9b}
^2{\cal B}_Q^{kmj} &=& ^1\Lambda_B\ ^{12}{\cal A}^{Bkmj}_Q +
^2\Lambda_{AB}\ ^{22}{\cal A}^{ABkmj}_Q +
^3\Lambda_{AB}\ ^{32}{\cal A}^{ABkmj}_Q,
\eey
and the reduced dissipation inequality
\bee{L9c}
^1\Lambda_B\ ^1{\cal C}^B + ^2\Lambda_{AB}\ ^2{\cal C}^{AB} +
^3\Lambda_{AB}\ ^3{\cal C}^{AB}\ \geq\ {\cal D}.
\vspace{.3cm}\ee
According to \R{D7} to \R{D9}, $^{1\bullet}{\cal A}$ belongs to 
the energy-momentum balance, $^{2\bullet}{\cal A}$ to the spin balance and
$^{3\bullet}{\cal A}$ to the gravitational field equations. The same is valid
for the $^\bullet{\cal C}$. This division allows a classification into
special cases which some of them are discussed in sect.\ref{EX}.

\subsection{Constraints of state space variables}

The particle number balance \R{1y}
\bee{L8x}
N^A\ab{A}\ =\ 0\ =\ (nu^A)\ab{A}\ =\ n\ab{A}u^A + nu^A\ab{A}
\ee
and also \R{10y} represent according to \R{12s1} a constraint for the state
space variables. How to handle these constraints with respect to the Liu
procedure, is expressed in the following  
\vspace{.3cm}\newline
$\blacksquare$ {\sf Proposition}\C{DipHH,HMPR,HRMP}: Constraints of the state
space variables do not influence the
Liu procedure, that means, the results achieved by the Liu procedure are
independent of introducing the constraints before or after performing it.$\!$ 
$\blacksquare$
\vspace{.3cm}\newline
Consequently, the particle number balance and the normalization of the
4-velocity do not generate additional constitutive restrictions.

\section{Special Cases\label{EX}}

\subsection{Ignoring the field equations}

The Mathisson-Papapetrou equations \R{v7y} and \R{u7y} are compatible with
Einsteins field equations \R{q7y}. Ignoring the field equations means that the
influence of the material on the geometry is not taken into account: the
curvature in \R{v7y} is given ad-hoc, but compatible with the field equations.
Decoupling the field equation means, we have to ignore in \R{L8} and \R{L9}
all quantities having the index combination $^{3\bullet}\circledcirc$. 
The Liu equations \R{L9a} and \R{L9b} become
\byy{L9d} 
^1\st{\Box}{{\cal B}}_k^{mj} &=& ^1\st{\Box}{\Lambda}_B\
^{11}{\cal A}^{Bmj}_k +
^2\st{\Box}{\Lambda}_{AB}\ ^{21}{\cal A}^{ABmj}_k ,
\\ \label{L9e}
^2\st{\Box}{{\cal B}}_Q^{kmj} &=& ^1\st{\Box}{\Lambda}_B\
^{12}{\cal A}^{Bkmj}_Q +
^2\st{\Box}{\Lambda}_{AB}\ ^{22}{\cal A}^{ABkmj}_Q ,
\eey
and the reduced dissipation inequality is
\bee{L9f}
^1\st{\Box}{\Lambda}_B\ ^1{\cal C}^B + ^2\st{\Box}{\Lambda}_{AB}\
^2{\cal C}^{AB}\ \geq\ \st{\Box}{{\cal D}}.
\vspace{.3cm}\ee
The constraints on the entropy --\R{L9d} and \R{L9e}-- and on the entropy
production \R{L9f} are changed with respect to the case, that the field
equations are taken into consideration --\R{L9a} to \R{L9c}. Ignoring the
field equations results
in altered material properties, if the Second Law is taken into account by
the Liu procedure.

\subsection{Vanishing spin}

If the spin tensor is set to zero, we obtain from \R{z7y}, \R{w7y} and \R{q7y}
\bee{L10}
T^{[AB]}\ =\ 0,\qquad T^{(AB)}\ab{A}\ =\ 0,\qquad \frac{2}{\kappa}R^{AB}
\ =\ 2 T^{(AB)} - \eta^{AB}T^D_D. 
\ee
The spin balance is satisfied by \R{L10}$_1$ and \R{L10}$_2$. Consequently, we
ignore all quantities with the index combination $^{2\bullet}\circledcirc$.
The energy-momentum balance is \R{D6} with the restriction that only the 
symmetric part ($AB\rightarrow (AB)$) comes into play and that $S^{CDE}$ is set
to zero. Thus, $^{11}{\cal A}$, $^{12}{\cal A}$ and $^{1}{\cal C}$ can be read
off from \R{D6}.
\vspace{.3cm}\newline
The gravitational field equations \R{L10}$_3$ become with \R{a25z}
\bee{L10a}
\frac{2}{\kappa}G^{Dkmj}_{ADBQ}e^Q_k{\ko{mj}}\ =\ -H^{D}_{ADB} +
2 T_{(AB)} - \eta_{AB}T^D_D.
\ee
To take into consideration the Second Law,
the following matrix elements vanish for state spaces of first order
\bee{L14} 
^{21}{\cal A},\ ^{22}{\cal A},\ ^2{\cal C},\ ^{31}{\cal A}\qquad
\mbox{are zero},
\ee
and we obtain
\byy{L14a}
^{32}{\cal A}_{ABQ}^{kmj} &=& \frac{2}{\kappa}G^{Dkmj}_{ADBQ}
\\ \label{L14b}
^3{\cal C}_{AB} &=& -H^{D}_{ADB} + 2 T_{(AB)} - \eta_{AB}T^D_D.
\eey
Consequently, the Liu equations \R{L9a} and \R{L9b} become
\byy{L14b} 
^1{\cal B}_k^{mj} &=& ^1\Lambda_B\ ^{11}{\cal A}^{Bmj}_k,
\\ \label{L14c}
^2{\cal B}_Q^{kmj} &=& ^1\Lambda_B\ ^{12}{\cal A}^{Bkmj}_Q +
^3\Lambda_{AB}\ ^{32}{\cal A}^{ABkmj}_Q,
\eey
and the reduced dissipation inequality is
\bee{L14d}
^1\Lambda_B\ ^1{\cal C}^B +
^3\Lambda_{AB}\ ^3{\cal C}^{AB}\ \geq\ {\cal D}.
\vspace{.3cm}\ee
This example of the Mathisson-Papapetrou equations demonstrates again that the
dissipation inequality and its exploitation have to be supplemented,
thus taking the Second Law into account. Even in the easy case of spinless
material, the Second Law cannot be ignored independently of taking the
gra\-vi\-tational field equations in the course of the Liu procedure into
account or not.

\subsection{Curvature insensitive material}

A material is called {\em curvature insensitive}, if its material
properties do not depend on $e^Q_k{\ko{m}}$. In that case, the
material does not contribute to the curvature, because the last term vanishes
in \R{D5} and with it the influence of the material upon the curvature.
Of course, the type of curvature sensitive material appear, always if the
curvature comes into play.
\vspace{.3cm}\newline
The curvature does not explicitly appear in the spin balance \R{u7y} and in the
dissipation inequality \R{3y} or \R{D3}. Consequently, the matrix elements
$^{22}{\cal A}$ and $^{2}{\cal B}$ vanish identically for curvature insensitive
material according to \R{D12} and \R{D13}. According to \R{L9b}, we obtain
\bee{L15}
0\ =\ ^1\Lambda_B\ ^{12}{\cal A}^{Bkmj}_Q +
^3\Lambda_{AB}\ ^{32}{\cal A}^{ABkmj}_Q,
\ee
an expression which also for curvature insensitive materials depends on the
curvature because the curvature appears explicitly
in the energy-momentum balance \R{v7y} and in the field
equations \R{t7y}.
\vspace{.3cm}\newline
The entropy production \R{L9c} may be influenced by the curvature: if the
$\Lambda$ depend on the curvature, the entropy production depends on it, too.

\section{Discussion}

The balance equations of continuum physics are formulated without taking a
special material into consideration, that means, one needs constitutive
equations to obtain by them a system of differential equations which is not
underdetermined. Constitutive equations are not arbitrary: they must be
modeled in such a way, that the solution of the resulting system of
differential equations satisfies the so-called material axioms. Among other
axioms, the Second Law may be the most prominent one. In continuum physics, it
runs as follows: Constitutive equations must have the property that the
entropy production is never negative for all
times and positions. A tool for finding such suitable constitutive equations
is the Liu procedure which is here performed in the framework of General
Relativity Theory. 
\vspace{.3cm}\newline
The characteristic of general-relativistic continuum physics is, that beyond
the balance equations, Einstein's gravitational field equations have to be
taken into ccount. In accordance with Maugin's view described in
footnote \ref{foot4}, we formally consider Einstein's gravitational equations
on the same footing as the balance equations of continuum thermodynamics,
rendering to apply the Liu procedure for the system of equations consisting of
the energy-momentum and spin balances --\R{2y} and \R{2ya}-- together with
Einstein's equations \R{4y}. 
\vspace{.3cm}\newline
Compatibility of the balances with the field equations enforces, that the
sources of the energy-momentum balance and of the spin balance
are not arbitrary, but result in the Mathisson-Papapetrou equations,
independently of including the field equations into the Liu procedure or not. 
As expected, the constraints for the constitutive equations are different, if
the gravitational field equations are included into the Liu procedure or are
ignored. Two other special cases are shortly discussed:
the spin-less system in General Relativity and curvature insensitive materials
which are the only materials in Special Relativity, but a special case in GRT.
\vspace{.3cm}\newline
In the present paper, the Liu procedure is performed for material mappings
defined on a first order state space. This results in the Liu conditions
--\R{L9a} and \R{L9b}-- that must be satisfied in order to guarantee the
compatibility of the combined system of differential equations
--\R{2y}, \R{2ya} and \R{4y}-- with the Second Law of thermodynamics \R{3y}.
\vspace{.3cm}\newline
Due to the complicated structure of the Liu equations in GRT, it
is extremely difficult to arrive at general statements with regard to the
compatibility of the general-relativistic continuum mechanics with the
Second Law. Starting out with a constitutive equation by a substantiate guess,
the Liu conditions --\R{L9a} to \R{L9c}-- differentiate between
gravitational fields as solutions of Einstein's equations: there could exist
solutions for which the constraints on the constitutive equations can be
satisfied and such ones for which this cannot be achieved. Regarding the
compatibility of the solutions of Einstein's equations with continuum
thermodynamics as an essential physical requirement, the Liu conditions
do not qualify all solutions of the gravitational equations as to be physically
relevant. In this sense, the Liu conditions act as a censor for the
constitutive equations with respect to the Second Law. 
\vspace{.3cm}\newline
Because the covariant curvature tensor \R{a25z} cannot be formulated
covariantly
as a function of the state space variables (see: sect.\ref{CTTR}), consequently
the Liu procedure cannot be performed in a covariant manner. That is no
disadvantage bearing into mind, that special solutions of the gravitational
equations cannot generally be characterized covariantly, either.

\section{Appendix} 

\subsection*{Curvature tensor in tetrad representation\label{CTTR}}

Starting out with \R{F4a}$_1$, we obtain from \R{F22a} that the connexion with
respect to the metric is a function of the tetrads and their first partial
derivatives
\bey\nonumber
\Gamma^b_{mi}(e^F_{a,b},e^F_a)\ =\hspace{10.1cm}
\\ \label{C1}
=\ \frac{1}{2}\eta_{AB}\eta^{DE}\Big\{e^B_q (e^A_{m}{\ko{i}}+ e^A_{i}{\ko{m}})
+ e^A_m (e^B_{q}{\ko{i}}- e^B_{i}{\ko{q}}) + e^A_i( e^B_{q}{\ko{m}}- 
e^B_{m}{\ko{q}})\Big\}e^b_D e^q_E .
\eey
We obtain from \R{F22a} by partial differentiation
\bee{C2} 
\Gamma^b_{mi,n}\ =\ \frac{1}{2}(g_{mq}{\ko{in}} + g_{iq}{\ko{mn}} -
g_{mi}{\ko{qn}})g^{bq} + \Gamma^p_{mi}g_{pq}g^{bq}{\ko{n}},
\ee
resulting in
\bey\nonumber
\Gamma^b_{mi}{\ko{n}}-\Gamma^b_{mn}{\ko{i}}\ =\hspace{8.8cm}
\\ \label{C3}
=\ \frac{1}{2}(g_{iq}{\ko{mn}} - g_{mi}{\ko{qn}} - g_{nq}{\ko{mi}} +
g_{mn}{\ko{qi}})g^{bq} +
g_{pq}(\Gamma^p_{mi}g^{bq}{\ko{n}}-\Gamma^p_{mn}g^{bq}{\ko{i}}).
\eey
Taking \R{F4a}$_1$ and 
\bee{C4}
g_{mn}{\ko{qi}}\ =\ \eta_{AB}(e^A_{m}{\ko{qi}}e^B_n + e^A_{n}{\ko{qi}}e^B_m +
e^A_{m}{\ko{q}}e^B_{n}{\ko{i}} +e^A_{m}{\ko{i}}e^B_{n}{\ko{q}})
\ee
into account, we obtain that the curvature tensor depends linearly on the
second derivatives of the tetrads. Consequently, it has the form which is
given in \R{a25ab}. Some easy, but long-winded calculations result in
\bey\nonumber
E^{buvw}_{mniG} &=& \eta_{AB}\eta^{CD}e^b_C e^q_D \Big[
(\delta^w_ie^B_n - \delta^w_ne^B_i)(\delta^A_G \delta^u_m \delta^v_q -
\delta^A_G \delta^u_q \delta^v_m)+
\\ \label{C5}
&&+(\delta^v_n e^A_q - \delta^v_q e^A_m)(\delta^B_G \delta^u_i \delta^w_n -
\delta^B_G \delta^u_n \delta^w_i)\Big],
\\ \nonumber
F^b_{mni} &=& \Gamma^j_{mi}\Gamma^b_{nj} - \Gamma^j_{mn}\Gamma^b_{ij} +
g_{pq}(\Gamma^p_{mi}g^{bq}{\ko{n}}-\Gamma^p_{mn}g^{bq}{\ko{i}}) +
\\ \nonumber
&&+\eta_{AB}g^{bq}\Big[(e^A_{m}{\ko{q}}- e^A_{q}{\ko{m}})
(e^B_{i}{\ko{n}}- e^B_{n}{\ko{i}}) +
\\ \label{C6}
&&\hspace{1.5cm}+e^A_{m}{\ko{n}}e^B_{i}{\ko{q}} -e^A_{m}{\ko{i}}e^B_{n}{\ko{q}}
+ e^A_{q}{\ko{i}}e^B_{n}{\ko{m}}-e^A_{q}{\ko{n}}e^B_{i}{\ko{m}}\Big].
\hspace{.5cm}
\eey
The final result is generated by inserting \R{F4a}$_1$ and \R{C1} into \R{C6}.

\end{document}